# New principle for unpolarized wideband reflectors

**Manoj Niraula and Robert Magnusson***

Department of Electrical Engineering, University of Texas at Arlington, Box 19016, Arlington TX 76019, USA

*magnusson@uta.edu

**Abstract:** There is immense scientific interest in the properties of resonant thin films embroidered with periodic nanoscale features. This device class possesses considerable innovation potential. Accordingly, we report unpolarized broadband reflectors enabled by a serial arrangement of a pair of polarized subwavelength gratings. Optimized with numerical methods, our elemental gratings consist of a partially-etched crystalline-silicon film on a quartz substrate. The resulting reflectors exhibit extremely wide spectral reflection bands in one polarization. By arranging two such reflectors sequentially with orthogonal periodicities, there results an unpolarized spectral band that exceeds those of the individual polarized bands. In the experiments reported herein, we achieve zero-order reflectance exceeding 97% under unpolarized light incidence over a 500-nm-wide wavelength band in the near-infrared domain. Moreover, the resonant unpolarized broadband accommodates an ultra-high-reflection band spanning ~85 nm and exceeding 99.9% in efficiency. The elemental polarization-sensitive reflectors based on one-dimensional resonant gratings have simple design, robust performance, and are straightforward to fabricate. Hence, this technology is a promising alternative to traditional multilayer thin-film reflectors especially at longer wavelengths of light where multilayer deposition may be infeasible or impractical.

## 1. Introduction

Resonant broadband reflection implemented with dielectric subwavelength gratings is a functional basis for a host of applications, including low-loss mirrors [1,2], narrow-linewidth bandpass filters [3], beam-transforming surfaces [4], and polarizers [5,6]. Devices in this class are often designed with a single dielectric thin-film layer and may incorporate carefully-crafted grating architectures to meet specifications for a particular application. Comparatively, obtaining a similar performance using traditional distributed Bragg stacks requires perhaps ~10-100 quarter- and/or half-wave layers [7]. Multilayer thin-film-based broadband dielectric mirrors represent an established technology. Their commercial manufacturers deposit multilayer films with high precision in layer thickness and index of refraction. The reflectivity and bandwidth of a traditional Bragg stack depends on the number of quarter-wave layer pairs and their refractive-index contrast [7]. At longer wavelengths of light, in the mid-infrared and THz domains for instance, multiple thin-film deposition is often impractical on account of typically-slow deposition techniques and issues in maintaining consistent deposition conditions across long time spans. Resonant grating reflectors achieve superior performance by way of parametric optimization [1,8]. Indeed, single-layer subwavelength grating mirrors can easily be scaled to be operative in any wavelength regime assuming availability of suitable materials. Practical challenges in fabrication include the need for periodic patterning, control of etch or imprint processes to maintain grating features, and management of process-induced scattering centers.

Broadband light reflection from a single-layer subwavelength grating is an interesting, perhaps counterintuitive, phenomenon as the constituent materials are optically transparent in nature. This phenomenon is understood as a low-$Q$ broadband resonance effect driven by laterally-guided Bloch modes [1,8,9]. Each mode represents a reflection peak (transmission dip) in the spectrum. In high-index grating architectures, wideband reflection can occur when reflection peaks corresponding to multiple guided-modes share a common, high-reflection band over a wide wavelength range. In a recent theoretical study, Magnusson showed that broadband mirrors based on subwavelength gratings can support wide reflection bands with efficiencies exceeding 99.99% [1]. Experimental demonstration of high-efficiency broadband mirrors based on this device class has been largely limited to input-polarization-dependent reflection [2,10]. There has been no prior experimental demonstration of wideband, high-efficiency, polarization-independent mirrors, which are critical in many practical applications.

Here, we report comprehensive design, fabrication, and spectral characterization of broadband mirrors based on subwavelength silicon gratings. As the constituent individual 1D-grating reflectors are polarization dependent, we arrange two gratings in series with orthogonal periodicity to achieve broadband reflection for unpolarized light incidence. The individual grating reflectors are separated by a gap to avoid evanescent field coupling between them. Excellent theoretical and experimental results are obtained, verifying the feasibility of the fundamental idea.

## 2. Elemental reflector concept

The canonical device structure applied here consists of a thin layer of partially-etched crystalline silicon (c-Si) on a quartz substrate as shown in Fig. 1(a). The c-Si layer, of total thickness *t*, is partially etched to create a 1D grating. The c-Si film lies on the *x*-*y* plane with the grating grooves parallel to the *y*-axis. The grating is periodic along the *x*-axis with a constant period of Λ. Each grating ridge has a height of $d_G$ and a width of *f*Λ, where *f* is the grating fill factor. We call this variety a "zero-contrast grating" (ZCG) as the grating ridges are matched to a sublayer made out of the same material; hence, no phase changes occur for a ridge mode transiting across the ridge/sublayer interface [1]. Using the particle-swarm optimization (PSO) method [9,11], we optimize the geometrical parameters of the grating structure to support high-efficiency broadband reflection for transverse-magnetic (TM) polarized light at normal incidence. Input light is in the TM (as opposed to transverse-electric, TE) polarization state if the magnetic (electric for TE) field vector is oriented along the grating grooves (*y*-axis). In the PSO calculations, we use an algorithm grounded in rigorous coupled-wave analysis (RCWA) [12] as the forward kernel to compute the spectral response. In our simulations, we account for the material dispersion in c-Si (refractive index = $n_H$, extinction coefficient = $k_H$) and quartz (refractive index = $n_S$). The $n_H$ and $k_H$ values for wavelengths up to 1450 nm are obtained from [13]. For longer wavelengths, $k_H$ is set to 0 and $n_H$ values reported in [14] are used. We calculate a dense grid of these optical constants through interpolation.

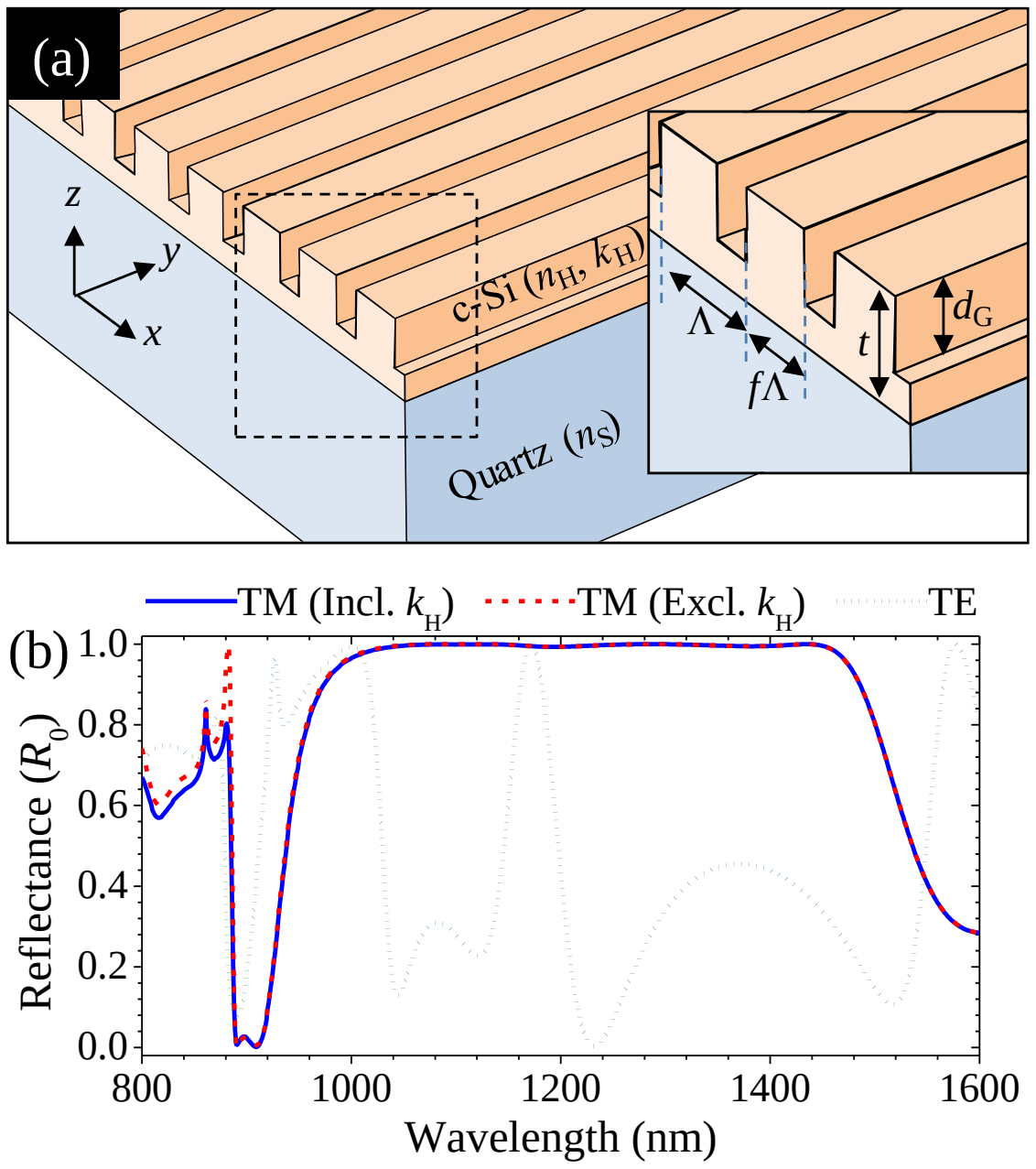

Fig. 1. Structure and performance of a subwavelength broadband mirror. (a) The device structure of a thin-layer of partially-etched c-Si on a quartz substrate. PSO optimized geometrical parameters are: period Λ = 560 nm, fill factor *f* = 0.66, c-Si layer thickness *t* = 520 nm, and grating depth $d_G$ = 330 nm. Refractive indices are $n_H$ and $n_S$ for c-Si and quartz, respectively. c-Si is modeled with extinction coefficient $k_H$. (b) Computed zero-order reflectance ($R_0$) of the device for light incidence at normal angle. Spectral response for TM polarization is presented for both lossy (solid blue) and lossless (dashed red) cases. The TE spectrum (dotted green) does not exhibit wideband reflection.

The PSO-optimized geometrical parameters are period Λ = 560 nm, fill factor *f* = 0.66, grating depth ($d_G$) = 332 nm, and film thickness *t* = 520 nm. Simulated zero-order reflectance ($R_0$) spectra for an elemental reflector with these geometrical parameters are shown in Fig. 1(b). The TM spectral response encompasses a 430-nm-wide spectral band extending from 1030 nm to 1460 nm with $R_0$ exceeding 99%. As the broadband reflection stems from resonances with low *Q* factors, we do not expect a significant drop in reflection efficiency due to extinction loss in the c-Si grating. Indeed, this is true as shown by the comparison between spectral responses for lossy and lossless cases in Fig. 1(b). A small degradation in reflection efficiency due to absorption loss is observed at wavelengths $\lambda \leq 1100$ nm. At $\lambda > 1100$ nm, the effect is negligible as c-Si becomes virtually lossless at that point.

## 3. Serial configuration for unpolarized reflection

It is well known that the 1D nature of these gratings results in a different spectral response for input TE and TM polarizations. This is clearly seen in Fig. 1(b). Taking advantage of this difference, broadband polarizers where one polarization is nearly completely transmitted and the other is nearly fully reflected have been proposed [6]. However, in many applications for broadband mirrors, input-polarization-independent performance is required. Recently, resonant reflectors that are periodic in 2D were proposed to reflect both TE and TM polarizations of light [15-17]. In contrast, here we demonstrate a polarization-independent spectral response by arranging two polarization-dependent ZCG-type reflectors such as those presented in Fig. 1 in series. This idea is illustrated in Fig. 2. There, two mirrors, denoted Device 1 and Device 2, have their grating grooves orthogonal to one another. For unpolarized light incidence, Device 1 nearly completely reflects the TM component of the light and partially transmits the TE component to Device 2. This transmitted light is now TM polarized with respect to Device 2 and is therefore almost completely reflected, resulting in near-complete reflection of both TE and TM components of the incident light.

The device concept in Fig. 2 relies on serial ZCG gratings that are spaced sufficiently far apart to avoid coupling between them via evanescent local fields. This concept differs from a theoretical proposal by Zhao *et al.* where cross-stacked gratings forming a two-layer woodpile photonic crystal are laid on a single substrate [18,19]. There, unpolarized light reflection across ~200 nm is a result spectral of overlap between reflective, but different, response for TE and TM polarized light input. Fabrication of woodpile photonic crystals is considerably more challenging than the simple single-layer devices provided currently.

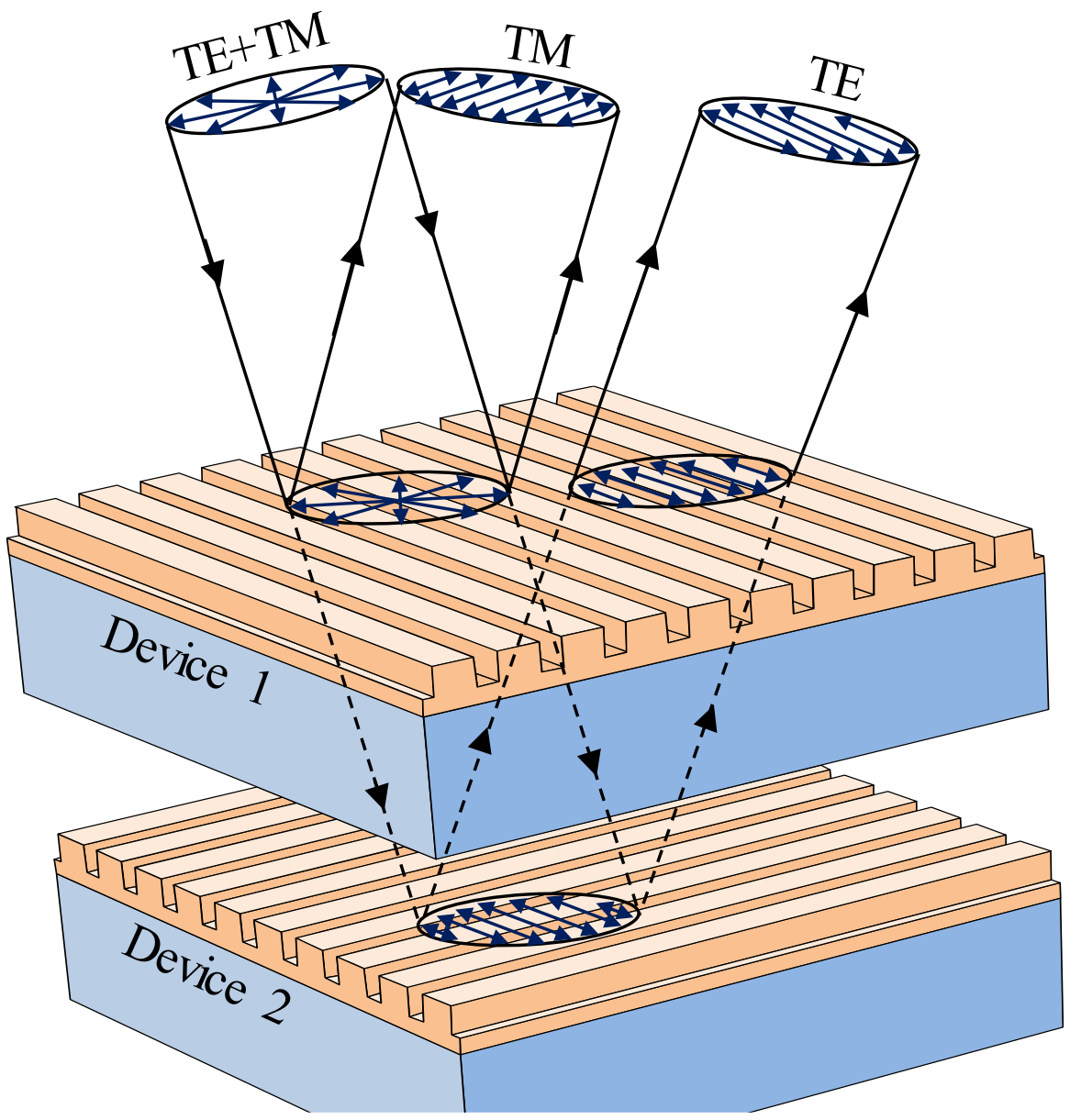

Fig. 2. Serial arrangement of elemental ZCG reflectors. The grating vectors of the two identical subwavelength grating mirrors, labeled as Device 1 and Device 2, are orthogonal. For unpolarized light incidence, Device 1 reflects the TM polarization components and Device 2 reflects the TE polarization components relative to Device 1. Note that the transmitted TE-polarized light from Device 1 is in the TM polarization state relative to Device 2.

## 4. Materials and methods

Our elemental ZCGs in Fig. 1(a) are fabricated using a commercially-available silicon-on-quartz (SOQ) wafer (Shin-Etsu Chemical Co). The SOQ wafer has a 520-nm-thick c-Si film on a quartz substrate. We use holographic lithography [20] to expose UVN-30, a negative photoresist (PR), creating a 1D mask. We reactive-ion etch (RIE) through the c-Si layer using $SF_6$ + $CF_4$ gas mixture. Residual PR after RIE is removed by ashing in an $O_2$ ambient. We characterize the fabricated devices using atomic force microscopy (AFM) and scanning electron microscopy (SEM) while optimizing the fabrication process to achieve the desired device geometry. Our ZCG reflectors for application in the serial configuration are fabricated using identical process conditions.

In the measurements, we use a low-coherence, super-continuum light source and a near-infrared optical spectrum analyzer. We measure the zero-order transmittance ($T_0$) of the elemental ZCG reflector and that of the sequential reflectors. For an elemental ZCG reflector, we measure $T_0$ as the transmitted signal normalized by the input signal for a particular polarization state. For the serial arrangement, two ZCGs are stacked with adjacent substrates; each substrate is 0.7 mm thick. In these experiments, the reference signal is the direct measured source spectrum with the sample removed and no other experimental components changed. Due to the subwavelength nature of the gratings, only zero-order propagating waves exist at wavelengths longer than the Rayleigh wavelength defined here in terms of the substrate index as $\lambda_R = \Lambda n_S$ (~850 nm). It follows that $R_0$ can be closely approximated as $1 - T_0$, especially at $\lambda > 1100$ nm where loss due to the absorption in silicon is negligible.

To characterize high-efficiency mirrors and verify their actual performance, direct measurement of the reflectance, as opposed to approximating it as $1 - T_0$, is typically desired. However, establishing a reliable and well-defined reference signal is challenging especially in multicomponent experiments where motion of a component in the beam path may lead to beam misalignment and associated errors. In order to verify the data collected by the transmission approach, we perform reflection experiments comparing directly the reflectance of a sample and a metal mirror. Thus, to avoid misalignment, we deposit, through vacuum sputtering, a gold film on a non-device area of the SOQ wafer. The ~100-nm-thick gold film serves as a reference mirror, enabling the measurement of a reference signal without having to replace the sample thereby preserving alignment. We measure the reflection intensities of the gold coating and an elemental ZCG reflector for TM polarized light input. Gold coatings are highly reflective above 900 nm in wavelength with reflectivity greater than 98% [21].

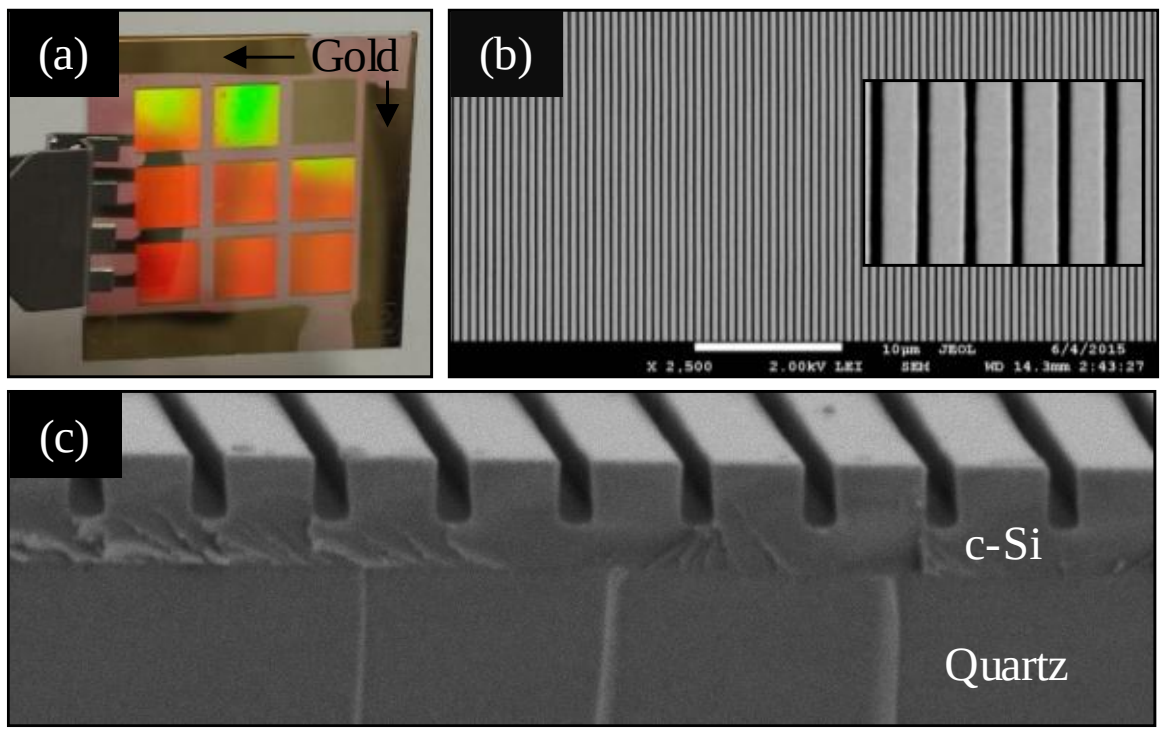

Fig. 3. Fabricated broadband mirror. (a) Photograph of nine fabricated devices on a 1x1 inch$^2$ SOQ wafer. Each device is 5x5 mm$^2$. Approximately 100-nm-thick gold film is sputtered on the non-device edges of the SOQ wafer. Scanning-electron micrographs showing (b) top-view and (c) cross-sectional images of a representative device.

## 5. Experimental results

A photograph of an array of nine fabricated devices on a single SOQ wafer is shown in Fig. 3(a). Here, we note that an array of broadband mirrors, each with different spectral characteristics and central wavelengths within the constraint of constant thickness, can be integrated on a single chip by varying the period and fill factor during holographic exposure; this is not generally feasible with multilayer Bragg stacks. Top-view SEM image of a representative device in Fig. 3(b) shows highly uniform grating lines. A cross-sectional SEM is shown in Fig. 3(c). Owing to its smoothness, we expect the scattering loss due to the fabrication process to be negligible. The ZCG mirrors fabricated for serial arrangement have similar geometrical parameters which are $f = 0.63$, $d_G = 330$ nm, $t = 520$ nm, and $\Lambda = 560$ nm.

Experimental $T_0$ and $R_0$ spectra of an elemental ZCG reflector for TM-polarized light incidence are shown in Fig. 4(a). A normal angle of incidence is maintained. The measured reflectance spectrum features a 490-nm-wide spectral band stretching from ~960 nm to ~1450 nm with $R_0$>0.97. The RCWA simulated TM spectrum in Fig. 4(b) shows an excellent *quantitative* agreement with the experimental spectrum in Fig. 4(a). In the simulations we use the geometrical device parameters obtained from AFM and SEM measurements of the actual samples. In addition, Figs. 4(c) and 4(d) show good *quantitative* agreement between measured and simulated spectra for TE polarized light incidence.

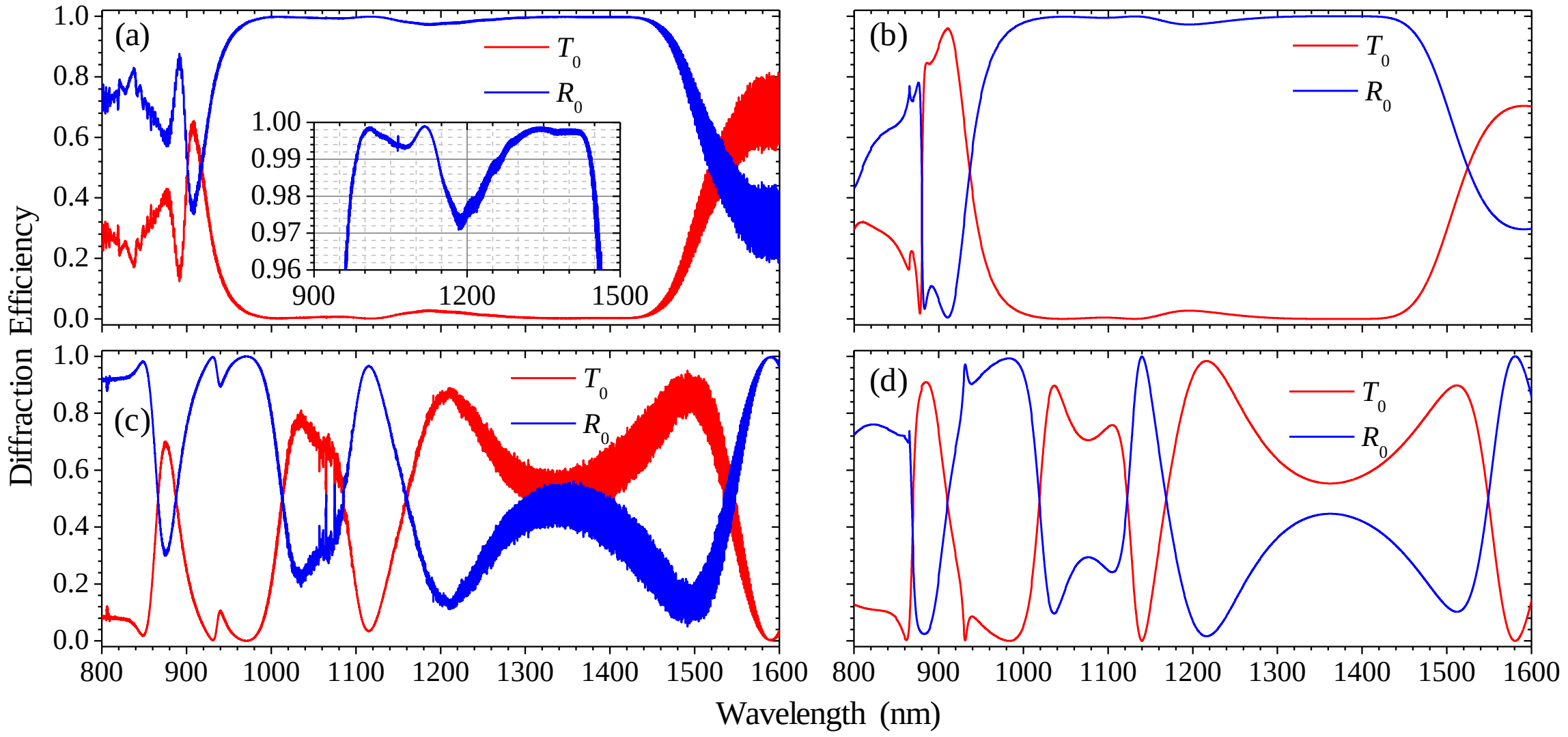

Fig. 4. Input-polarization-dependent spectral response of an elemental ZCG reflector. (a) Experimental and (b) simulated $T_0$ and $R_0$ spectra for TM polarized light incidence. (c) Experimental and (d) simulated $T_0$ and $R_0$ spectra for TE-polarized light incidence. Device parameters are fill factor $f$ = 0.63, grating depth $d_G$ = 330 nm, c-Si film thickness $t$ = 520 nm, and grating period $\Lambda$ = 560 nm. Lossy c-Si films and an infinitely-thick quartz substrate are assumed in (b) and (d). The inset in (a) shows the wavelength band for $R_0 > 96\%$. In the experiments reported in (a) and (c), $R_0$ is approximated as $1 - T_0$. Spectral noise in the measurements at $\lambda > 1200$ nm is due to Fabry-Pérot resonance effects in an optical-fiber segment used in the data collection.

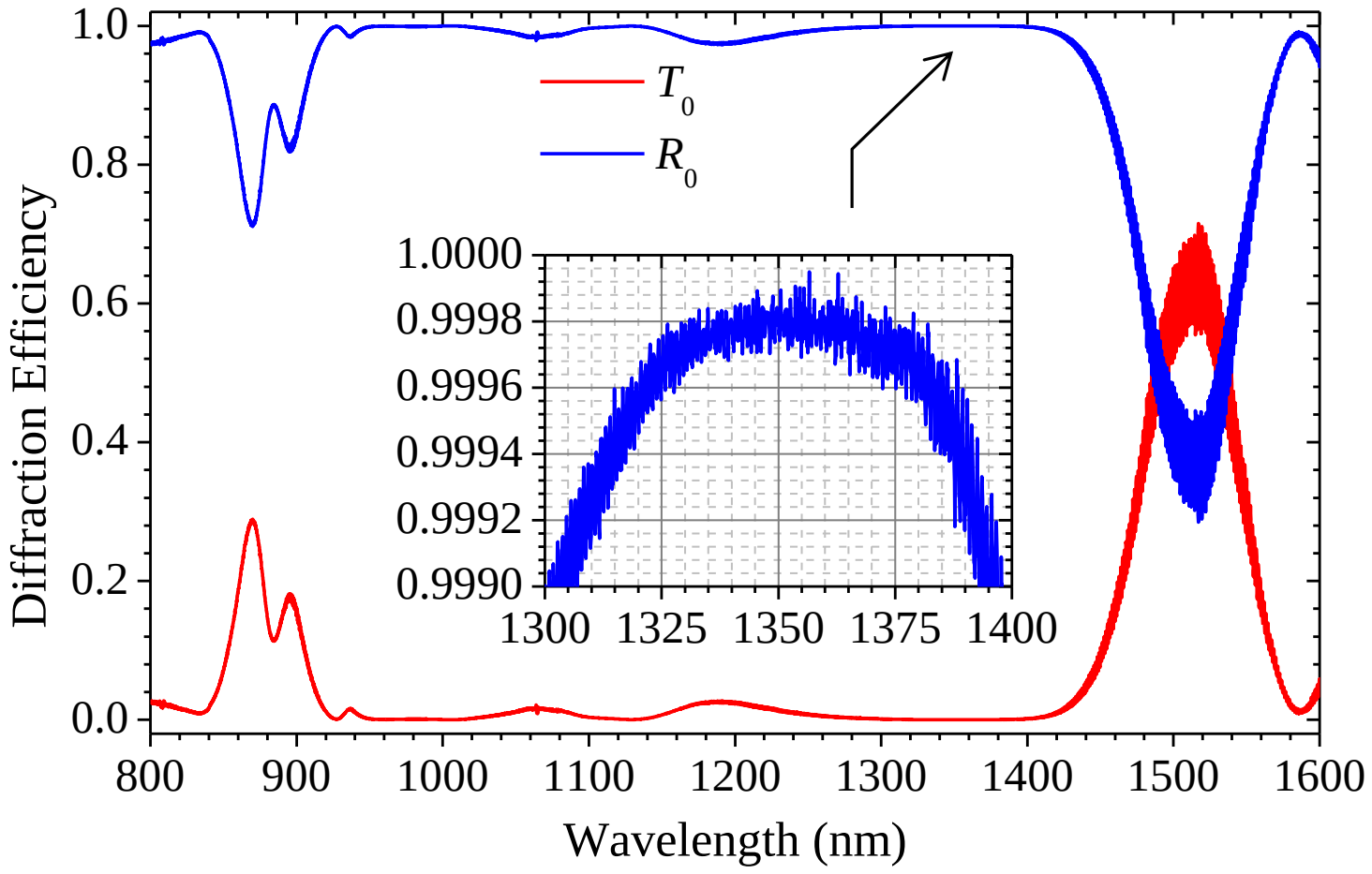

Fig. 5. Polarization-independent spectral response of two serially-arranged ZCGs. Measured $T_0$ and $R_0$ data using the setup illustrated in Fig. 2 for unpolarized light at normal incidence. $R_0$ is approximated as $1 - T_0$. The inset shows a detailed view of the wavelength band with $R_0 > 99.9\%$.

Outstanding experimental performance of an elemental ZCG reflector is limited to TM polarization, as is evident by the experimental results in Fig. 4. To demonstrate unpolarized broadband reflection, we concatenate two ZCGs as in Fig. 2 and proceed to experimental characterization. Measurement results using unpolarized light at normal incidence are shown in Fig. 5. Here, the 520-nm-wide reflection band from ~915 nm to ~1435 nm lies above 97% in reflectance. Moreover, an 85-nm-wide reflection band centered at $\lambda$ = 1350 nm has ultra-high-reflectance with efficiency exceeding 99.9% and parts of that subband near 99.99%. We note that experimental performance of this class of devices is limited only by fabrication imperfections as the theoretical performance of an elemental ZCG reflector can exceed 99.99% for wide spectral bands [1].

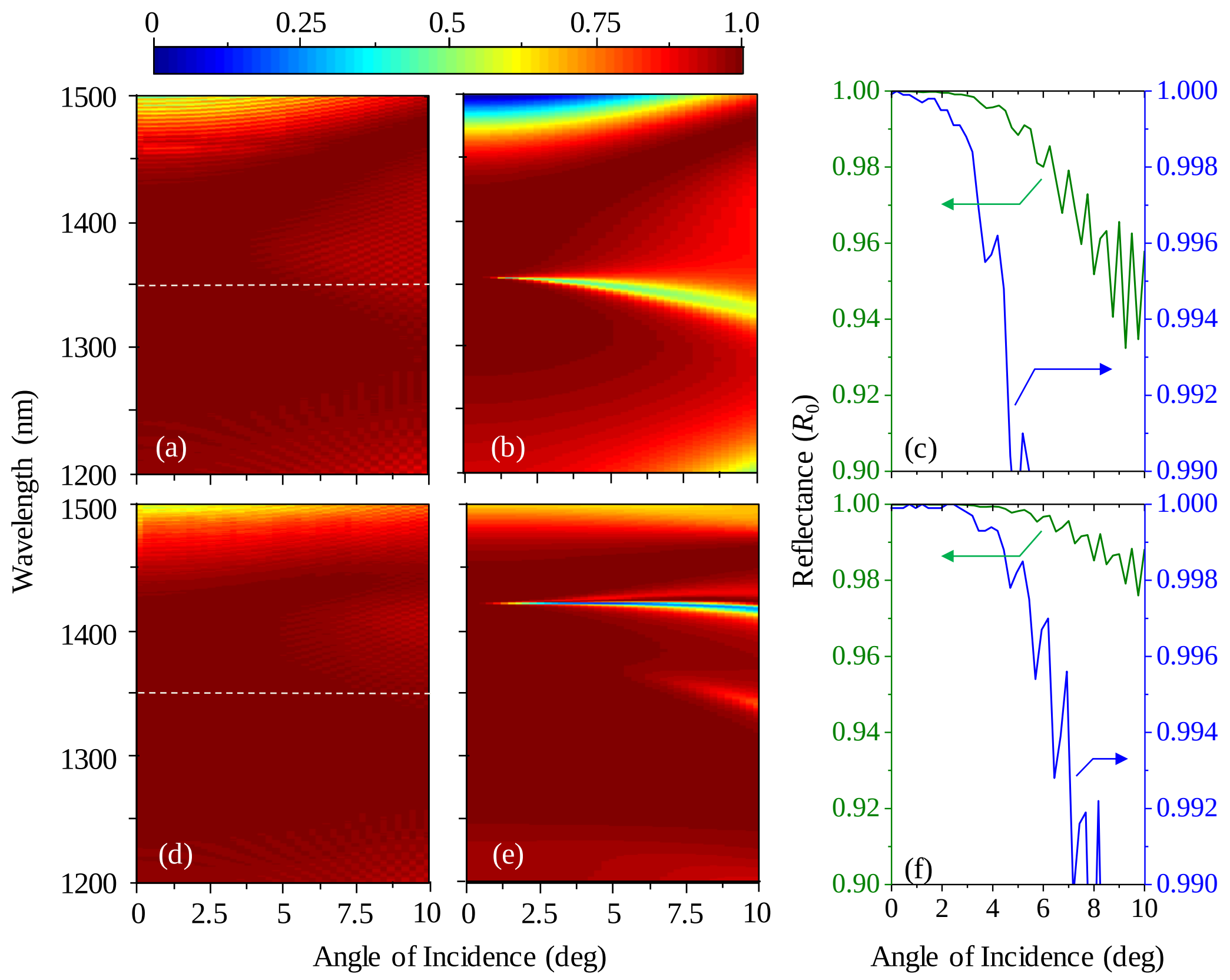

Fig. 6. A study of the angular response. (a) Measured and (b) simulated $R_0$ maps for different polar angles corresponding to an elemental ZCG mirror for TM-polarized light incidence. (c) Angular $R_0$ corresponding to $\lambda$ = 1350 nm in (a). (d)-(e) $R_0$ maps for serial ZCG mirrors for unpolarized light incidence showing (d) measurement and (e) simulation results. (f) Angular $R_0$ corresponding to $\lambda$ = 1350 nm in (d). In the measurements, $R_0$ is approximated as $1 - T_0$. Dashed lines in (a) and (d) represent $\lambda$ = 1350 nm.

The experimental spectra in Figs. 4(a) and 5 are fairly robust relative to small deviations in the angle of incidence. To quantify this, we set the input wavelength to 1350 nm and assess the behavior of the surrounding reflection band with extraordinarily high efficiency for small variations in the polar angle of incidence, $\theta$. For a ZCG mirror, as illustrated in Fig. 1(a), we assume that the incident wave vector remains in the x-z plane and $\theta$ is defined as the angle between the wave vector and the z-axis. Results from this study are presented in Fig. 6. Measured and simulated angle-dependent reflectance maps corresponding to a polarization-dependent ZCG under TM-polarized light incidence are shown in Figs. 6(a) and 6(b), respectively. Similar maps for unpolarized light corresponding to serial devices as in Fig. 2 are shown in Fig. 6(d) for measurement and Fig. 6(e) for simulation. Figures 6(c) and 6(f) show experimental data of $R_0$ vs. $\theta$ at $\lambda$ = 1350 nm corresponding to Figs. 6(a) and 6(d), respectively. These experimental results show robust performance of both types of reflectors for small angular variations. For instance, at $\lambda$ = 1350 nm, $R_0 \geq 99.9\%$ is sustained for $\theta \leq 2.75°$ for the polarized ZCG reflector whereas this value of $R_0$ maintains for $\theta \leq 4.25°$ for the unpolarized configuration. At the same wavelength, $R_0 \geq 99\%$ covers up to 4.75° and 7.75° in $\theta$ for these polarization-dependent and polarization-independent cases, respectively. The deviation between theory and experiment in Fig. 6 derives, at least partly, from the fact that the simulation applies ideal, completely coherent plane waves whereas the experiment employs an incoherent Gaussian beam.

Finally, measured reflection intensity of the sputtered gold mirror in comparison with the reflectance of one of our ZCG reflectors is shown in Fig. 7. TM polarization and a normal angle of incidence are employed. Here, for ~480-nm-wide wavelength range from ~960 nm to ~1440 nm, the measured reflectance of the grating device is generally on par or higher than that of the gold mirror. This implies that device reflectivity at the marked ~480-nm-wide band is ≥98% as claimed above. This is consistent with the experimental results presented in Fig. 4(a). This supports our use of the approach taken in this research that $R_0 = 1 - T_0$ is a reliable way to quantify our devices.

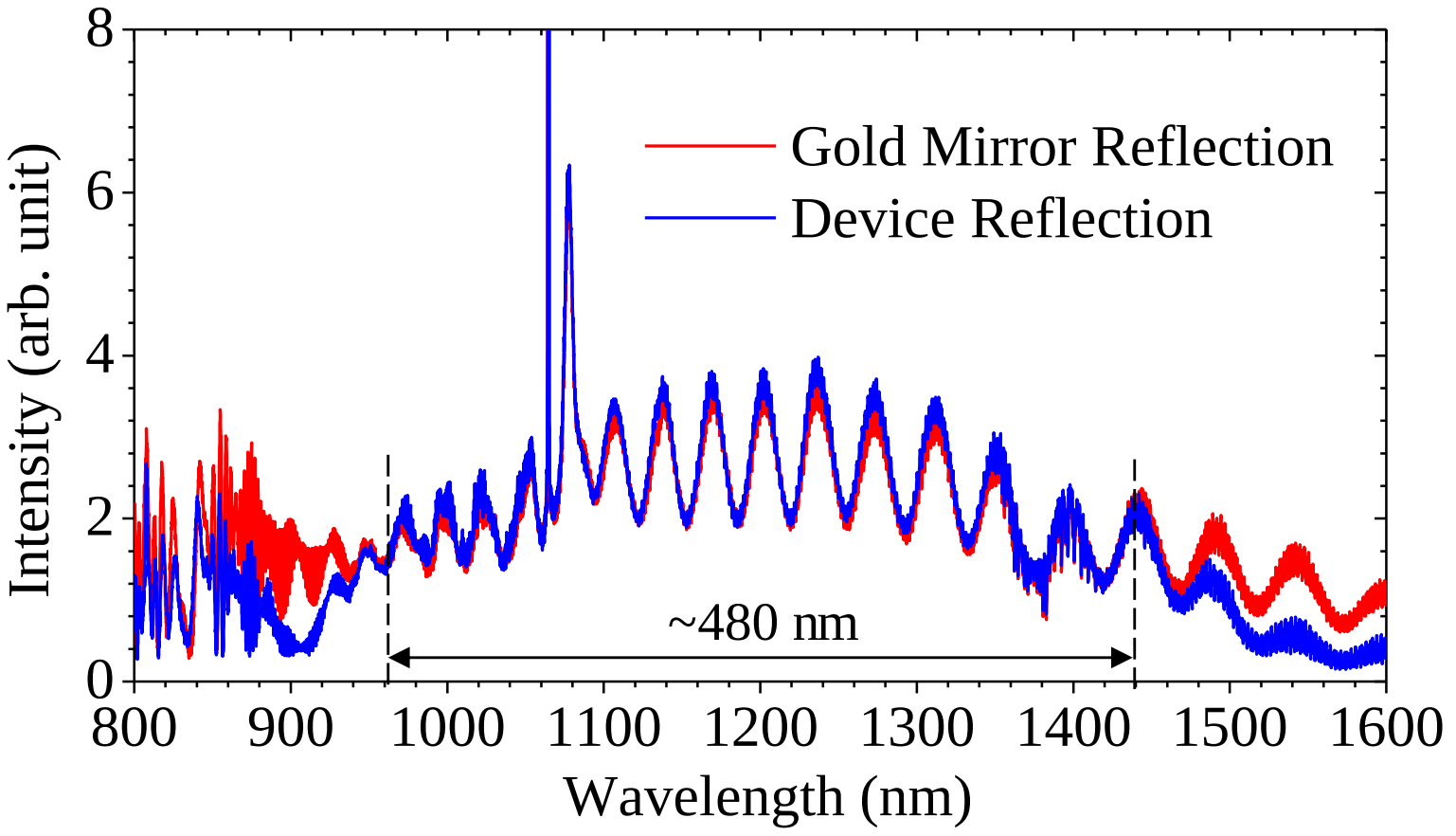

Fig. 7. Reflection comparison with a gold mirror. Measured reflection intensities using a gold mirror (red line) and an elemental ZCG mirror (blue line) for TM-polarized light at normal incidence are shown. For a ~480-nm-wide spectral band centered at ~1230 nm, the measured intensity for the ZCG device is generally higher than the measured intensity for the gold mirror. The intensity spike at $\lambda \approx 1060$ nm is characteristic of the light source used in the measurement.

## 7. Conclusions

In summary, we report successful design, fabrication, and characterization of unpolarized high-efficiency broadband mirrors based on subwavelength c-Si gratings. Our sequential arrangement concept is fundamentally different from coupled woodpile gratings and 2D photonic crystals previously reported in the literature. In fact, our unpolarized reflectors possess the extremely wide polarization-dependent reflection bands obtainable with optimized 1D ZCGs [1]. The relative bandwidth, defined as the ratio of unpolarized-reflection bandwidth ($\Delta\lambda$) for $R_0$>97% and center wavelength ($\lambda_C$), in our experiments, is $\Delta\lambda/\lambda_C$ ~44%. For comparison, the relative unpolarized-reflection bandwidths for the woodpile structure designed by Zhao *et al.* [18] and the 2D ZCG grating structure proposed by Ko *et al.* [17] are ~14% and ~24%, respectively. Generally, in 2D, or coupled 1D, resonant reflectors, the unpolarized reflection bandwidth is limited by a less-reflective polarization state. In contrast, a fractional band with high TE reflectance adjacent to the wide TM band provided by the ZCG reflectors reported here actually extends the unpolarized bandwidth. This is shown by our numerical and experimental results. In design and optimization of reflectors belonging to the device class presented herein, bandwidth enhancement by such means should be considered.

In spite of the conceptual simplicity of the ideas presented herein, their practical significance is potentially enormous. The basic polarized ZCG reflectors have simple design, robust performance, and are straightforward to fabricate. Hence, this technology may be a promising alternative to multilayer thin-film reflectors particularly at longer wavelengths of light where film deposition may be infeasible or impractical.

## Acknowledgements

This research was supported by NSF Award No: IIP-1444922. The authors thank Shin-Etsu Chemical Co, Ltd., Japan for providing the SOQ wafers.